\documentstyle[12pt]{article}

\def\beqa{\begin{eqnarray}}
\def\eeqa{\end{eqnarray}}
\def\beq{\begin{equation}}
\def\eeq{\end{equation}}
\def\beqa{\begin{eqnarray}}
\def\eeqa{\end{eqnarray}}
\newcommand{\be}{\begin{equation}}
\newcommand{\eq}{\end{equation}}
\newcommand{\ee}{\end{equation}}
\newcommand{\ba}{\begin{array}}
\newcommand{\ea}{\end{array}}
\input{tcilatex}

\begin{document}

\begin{flushright}
CAMS/99-05\\
hep-th/9911195
\end{flushright}
\vspace{0.5cm}

\begin{center}
\baselineskip=16pt {\Large {\bf Magnetic Strings In Five Dimensional Gauged
Supergravity Theories}} \vskip1 cm {A. H. Chamseddine and \thinspace W. A.
Sabra }\footnote{%
e-mail: chams@aub.edu.lb, {ws00@aub.edu.lb}} \vskip1cm 
\centerline{{\em
Center for Advanced Mathematical Sciences (CAMS),} } \centerline{{\em and}} %
\centerline{{\em Physics Department, American University of Beirut, Lebanon}}
\end{center}

\vskip1 cm \centerline{\bf ABSTRACT} \vspace{-0.3cm}

\bigskip

Magnetic BPS string solutions preserving quarter of supersymmetry are
obtained for all abelian gauged $d=5$ $N=2$ supergravity theories coupled to
vector supermultiplets. Due to a ``generalised Dirac quantization''
condition satisfied by the minimized magnetic central charge, the string
metric takes a universal form for all five dimensional gauged theories. %
\vfill\eject

\section{Introduction}

Black hole solutions of gauged extended supergravities have recently been
the subject of intense research activities \cite{romans, bcs1, bcs2, klemm,
Duffliu, Sabra}. This is largely motivated by the newly proposed
correspondence between anti-de Sitter space, the ground state of gauged
supergravity, and conformal field theories on its boundary \cite{ads}.
Anti-de Sitter black hole solutions which break some or all of
supersymmetries may be of relevance to the proposed anti-de Sitter/Conformal
Field Theory correspondence.

BPS-saturated black holes constitute a large class of non-trivial
gravitational backgrounds preserving some part of the superymmetry of the
theory. Supersymmetric BPS solutions for the theory of $N=2$ gauged four
dimensional pure supergravity were first considered in \cite{romans}. Also,
BPS electrically charged solutions were found for the four dimensional $N=8$
and $N=2$ supergravity theories with vector multiplets in \cite{Duffliu,
Sabra}. In five dimensional gauged $N=2$ supergravity coupled to vector
supermultiplets\cite{gst1}, static spherically symmetric electrically
charged BPS-saturated black holes were considered in \cite{bcs1}. A common
feature of all these solutions is that they are supersymmetric and have
naked singularities. BPS-saturated topological black holes in gauged
supergravity, also with naked singularities, were obtained in \cite{klemm}.
Non-extreme electrically charged static black hole solutions of $N=2$ five
dimensional gauged supergravities were studied in \cite{bcs2}. The geometry
of these solutions, in particular their singularity structure, and the
domain of the parameters of the ADM mass for which horizons exist as well as
their thermodynamic features were also studied. Such features could
potentially provide an insight into dynamics of Yang-Mills theories with
broken supersymmetry.

BPS magnetic string solutions which break half of supersymmetry in the
theory of ungauged $N=2$ five-dimensional supergravities were constructed in 
\cite{magnetic}. These solutions correspond , in models which can be
obtained as compactifications of M-theory on a Calabi-Yau manifold, to five
branes wrapping around the homology cycles of the Calabi-Yau compact space.
Near the horizon, the supersymmetry of these solutions gets enhanced and
fully restored. The horizon geometry of these solutions is identified with
the space $AdS_{3}\times S^{2}$.

Our purpose in this paper is to study spherically symmetric magnetically
charged string solutions of five dimensional $N=2$ gauged supergravity
theory coupled to vector supermultiplets\cite{gst2}. We organize this work
as follows. In Section 2, a brief review of five dimensional $N=2$
supergravity is given within the context of very special geometry. In
Section 3, magnetic string solution which preserves 1/4 of the $N=2$
supersymmetry are explicitly derived. It it shown that for any choice of the
prepotential ${\cal V}$ which defines the five dimensional theory, the
string metric takes a universal form independent of the charge configuration
of the solution. This universality is a consequence of the fact that the
solutions depend on the magnetic central charge which, for the
supersymmetric configuration, has to satisfy a generalised ``Dirac
quantization condition''. As discussed for the ungauged cases \cite{magnetic}%
, a subclass of solutions of the $N=2$ models are also solutions for models
with more supersymmetry; $i.e.$, $N=4$ and $N=8$ supergravity. This is the
three-charge configuration with no self-intersections. The last section
includes a summary and a discussion.

\section{$D=5$ $\ N=2$ Gauged Supergravity}

The theory of five-dimensional $N=2$ supergravity coupled to abelian vector
supermultiplets can be obtained by compactifying eleven-dimensional
supergravity, the low-energy theory of M-theory, on a Calabi-Yau three-folds 
\cite{cy}. The massless spectrum of the theory contains $(h_{(1,1)}-1)$
vector multiplets with real scalar components, and thus $h_{(1,1)}$ vector
bosons (the additional vector boson is the graviphoton). The theory also
contains $h_{(2,1)}+1$ hypermultiplets, where $h_{(1,1)}$ and $h_{(2,1)},$
are the Calabi-Yau Hodge numbers. Gauged supergravity theories are obtained
by gauging a subgroup of the R-symmetry group; the automorphism group of the
supersymmetry algebra. The gauged $d=5,$ $N=2$ supergravity theories are
obtained by gauging the $U(1)$ subgroup of the $SU(2)$ automorphism group of
the supersymmmetry algebra \cite{gst2}. This is achieved by introducing a
linear combination of the abelian vector fields already present in the
ungauged theory, i.e. $A_{\mu }=V_{I}A_{\mu }^{I}$, with a coupling constant 
$g$. The coupling of the fermi-fields to the $U(1)$ vector field breaks
supersymmetry, and therefore gauge-invariant $g$-dependent terms must be
introduced in order to preserve $N=2$ supersymmetry. In a bosonic
background, this amounts to the addition of a $g^{2}$-dependent scalar
potential \ $V$\cite{gst2, bcs1}.

The bosonic part of the gauged supersymmetric $N=2$ Lagrangian which
describes the coupling of vector multiplets to supergravity is given by 
\begin{eqnarray}
e^{-1}{\cal {L}} &=&{\frac{1}{2}}R+g^{2}V-{\frac{1}{4}}G_{IJ}F_{\mu \nu
}{}^{I}F^{\mu \nu J}-{\frac{1}{2}}{g}_{ij}\partial _{\mu }\phi ^{i}\partial
^{\mu }\phi ^{j}  \nonumber \\
&&+{\frac{e^{-1}}{48}}\epsilon ^{\mu \nu \rho \sigma \lambda }C_{IJK}F_{\mu
\nu }^{I}F_{\rho \sigma }^{J}A_{\lambda }^{k}  \label{action}
\end{eqnarray}
with the space-time indices $(\mu ,\nu )=0,1,\cdots ,4$, $R$ is the scalar
curvature, $F_{\mu \nu }^{I}$ are the abelian field-strength tensors and $e=%
\sqrt{-g}$ is the determinant of the F\"{u}nfbein $e_{m}^{\ a}$, $V$ is the
potential given by 
\begin{equation}
V(X)=V_{I}V_{J}\left( 6X^{I}X^{J}-{\frac{9}{2}}{g}^{ij}\partial
_{i}X^{I}\partial _{j}X^{J}\right) ,
\end{equation}
where $X^{I}$ represent the real scalar fields which satisfy the following
condition 
\begin{equation}
{\cal V}={\frac{1}{6}}C_{IJK}X^{I}X^{J}X^{K}=1\ .
\end{equation}

The physical quantities in (\ref{action}) can all be expressed in terms of
the homogeneous cubic polynomial ${\cal V}$ which defines ``very special
geometry'' \cite{very}. We also have the relations 
\begin{eqnarray}
G_{IJ} &=&-{\frac{1}{2}}\partial _{I}\partial _{J}\log {\cal V}\Big|_{{\cal V%
}=1}  \nonumber \\
{g}_{ij} &=&\partial _{i}X^{I}\partial _{j}X^{J}G_{IJ}\Big|_{{\cal V}=1}\ ,
\end{eqnarray}
where $\partial _{i}$ and $\partial _{I}$ refer, respectively, to a partial
derivative with respect to the scalar field $\phi ^{i}$ and $%
X^{I}=X^{I}(\phi ^{i})$.

\bigskip It is worth pointing out that for Calabi-Yau compactification, $%
{\cal V}$ represents the intersection form, $X^{I}$ and $X_{I}={\frac{1}{6}}%
C_{IJK}X^{J}X^{K}$ correspond, respectively, to the size of the two- and
four-cycles and $C_{IJK}$ are the intersection numbers of the Calabi-Yau
threefold.

\section{Magnetic String Solutions}

In this section the BPS extended magnetic solutions preserving 1/4 of the $%
N=2$ supersymmetry are constructed. This is achieved by solving for the
vanishing of supersymmetry transformation of the gravitino and gauginos
fields in a bosonic background. The supersymmetry transformation of these \
fermionic fields in a bosonic background are given by \cite{bcs1} 
\begin{eqnarray}
\delta \psi _{\mu } &=&\left( {\cal {D}}_{\mu }+{\frac{i}{8}}X_{I}(\Gamma
_{\mu }{}^{\nu \rho }-4\delta _{\mu }{}^{\nu }\Gamma ^{\rho })F_{\nu \rho
}{}^{I}+{\frac{1}{2}}g\Gamma _{\mu }X^{I}V_{I}-{\frac{3i}{2}}gV_{I}A_{\mu
}^{I}\right) \epsilon ,  \nonumber \\
\delta \lambda _{i} &=&\left( {\frac{3}{8}}\Gamma ^{\mu \nu }F_{\mu \nu
}^{I}\partial _{i}X_{I}-{\frac{i}{2}}g_{ij}\Gamma ^{\mu }\partial _{\mu
}\phi ^{j}+{\frac{3i}{2}}gV_{I}\partial _{i}X^{I}\right) \epsilon  \label{st}
\end{eqnarray}
where $\epsilon $ is the supersymmetry parameter and ${\cal {D}_{\mu }}$ is
the covariant derivative\footnote{%
we use the metric $\eta ^{ab}=(-,+,+,+,+)$, $\{{\Gamma ^{a},\Gamma ^{b}}%
\}=2\eta ^{ab}$, ${\cal {D}}_{\mu }=\partial _{\mu }+{\ \frac{1}{4}}\omega
_{\mu ab}\Gamma ^{ab}$, $\omega _{\mu ab}$ is the spin connection, and $%
\Gamma ^{{\nu }}$ are Dirac matrices and $\Gamma ^{{a}_{1}{a}_{2}\cdots {%
a_{n}}}={\frac{1}{n!}}\Gamma ^{\lbrack {a_{1}}}\Gamma ^{{a_{2}}}\cdots
\Gamma ^{{a_{n}}]}$.}.

A general spherically symmetric string solution can be written in the
following form

\begin{equation}
ds^{2}=-e^{2V}dt^{2}+e^{2T}dz^{2}+e^{2U}dr^{2}+r^{2}\left( d\theta ^{2}+\sin
^{2}\theta d\phi ^{2}\right)
\end{equation}
and for the gauge fields we take 
\begin{eqnarray}
A_{\phi }^{I} &=&-q^{I}\cos \theta ,\qquad  \nonumber \\
F_{\theta \phi }^{I} &=&q^{I}\sin \theta ,  \label{gf}
\end{eqnarray}
where the functions $(U,V,T)$ are functions of $r$, and $(\theta ,\phi )$
are the polar coordinates of the 2-sphere.

\bigskip

The Funfbein of the above metric are given by

\begin{eqnarray*}
e_{t}^{0} &=&e^{V},\text{ \ \ \ \ \ \ \ \ \ \ \ \ \ }e_{t}^{0}=e^{-V} \\
\text{ }e_{z}^{1} &=&e^{T},\text{ \ \ \ \ \ \ \ \ \ \ \ \ \ }e_{1}^{z}=e^{-T}
\\
e_{r}^{2} &=&e^{U},\text{ \ \ \ \ \ \ \ \ \ \ \ \ }e_{2}^{r}=e^{-U} \\
e_{_{\theta }}^{3} &=&r\text{ ,\ \ \ \ \ \ \ \ \ \ }\ \ \ \ \text{\ }%
e_{3}^{\theta }=\frac{1}{r} \\
e_{_{\phi }}^{4} &=&r\text{ }\sin \theta ,\text{\ \ \ \ \ \ \ \ \ }%
e_{_{4}}^{\phi }=\frac{1}{r\sin \theta }.
\end{eqnarray*}

The spin connections for the above metric are given by

\begin{eqnarray}
w_{t02} &=&-V^{\prime }e^{V-U},  \nonumber \\
w_{z12} &=&T^{\prime }e^{T-U},  \nonumber \\
w_{\theta 23} &=&-e^{-U},  \nonumber \\
w_{\phi _{24}} &=&-e^{-U}\sin \theta ,  \nonumber \\
w_{\phi 34} &=&-\cos \theta .  \label{sss}
\end{eqnarray}

where $\left( 0,1,\text{ }2,\text{ }3,\text{ }4\right) $ represent the flat
indices.

\bigskip Then from the supersymmetry transformation of the fermionic fields
we obtain

\begin{eqnarray}
\delta \psi _{t} &=&\left( \partial _{t}+{\frac{1}{2}}V^{\prime
}e^{V-U}\Gamma _{02}+\frac{i}{4}Z\frac{e^{V}}{r^{2}}\Gamma _{034}+\frac{1}{2}%
ge^{V}X^{I}V_{I}\Gamma _{0}\right) \epsilon  \nonumber \\
\delta \psi _{z} &=&\left( \partial _{z}+{\frac{1}{2}}T^{\prime
}e^{T-U}\Gamma _{12}+\frac{i}{4}Z\frac{e^{T}}{r^{2}}\Gamma _{134}+\frac{1}{2}%
ge^{T}X^{I}V_{I}\Gamma _{1}\right) \epsilon  \nonumber \\
\delta \psi _{\theta } &=&\left( \partial _{\theta }-{\frac{1}{2}}%
e^{-U}\Gamma _{23}-\frac{i}{2}Z\frac{1}{r}\Gamma _{4}+\frac{1}{2}%
grX^{I}V_{I}\Gamma _{3}\right) \epsilon  \nonumber \\
\delta \psi _{\phi } &=&\left( \partial _{\phi }-({\frac{e^{-U}}{2}}\Gamma
_{24}-{\frac{iZ}{2r}}\Gamma _{3}-\frac{gr}{2}\Gamma _{4})\sin \theta +\frac{1%
}{2}(3igq^{I}V_{I}-\Gamma _{34})\cos \theta \right) \epsilon  \nonumber \\
\delta \psi _{r} &=&\left( \partial _{r}+{\frac{ie^{U}}{4}}\frac{Z}{r^{2}}%
\Gamma _{234}+\frac{1}{2}{ge}^{U}X^{I}V_{I}\Gamma _{2}\right) \epsilon
\end{eqnarray}

where $Z=q^{I}X_{I}$ is the magnetic central charge. As supersymmetric
breaking conditions we take the following conditions \ 
\begin{eqnarray}
\Gamma _{3}\Gamma _{4}\epsilon &=&i\epsilon ,  \nonumber \\
\Gamma _{2}\epsilon &=&-\epsilon .  \label{bc}
\end{eqnarray}
Then, the above transformations reduce to 
\begin{eqnarray}
\delta \psi _{t} &=&\left( \partial _{t}-{\frac{1}{2}}(V^{\prime }e^{V-U}+Z%
\frac{e^{V}}{2r^{2}}-ge^{V}X^{I}V_{I})\Gamma _{0}\right) \epsilon , 
\nonumber \\
\delta \psi _{z} &=&\left( \partial _{z}-{\frac{1}{2}}(T^{\prime }e^{T-U}+Z%
\frac{e^{T}}{2r^{2}}-ge^{T}X^{I}V_{I})\Gamma _{1}\right) \epsilon , 
\nonumber \\
\delta \psi _{\theta } &=&\left( \partial _{\theta }-{\frac{1}{2}}(e^{-U}-Z%
\frac{1}{r}-grX^{I}V_{I})\Gamma _{3}\right) \epsilon ,  \nonumber \\
\delta \psi _{\phi } &=&\left( \partial _{\phi }-{\frac{i}{2}}%
(1-3gV_{I}q^{I})\cos \theta -{\frac{1}{2}}(e^{-U}-Z\frac{1}{r}%
-grX^{I}V_{I})\sin \theta \Gamma _{4}\right) \epsilon ,  \nonumber \\
\delta \psi _{r} &=&\left( \partial _{r}+{\frac{e^{U}}{2}}(\frac{Z}{2r^{2}}-{%
g}X^{I}V_{I})\right) \epsilon .
\end{eqnarray}

\bigskip

The vanishing of the above equations implies the following conditions on the
supersymmetry spinor $\epsilon ,$ 
\begin{eqnarray}
\partial _{t}\epsilon &=&0,  \nonumber \\
\partial _{\theta }\epsilon &=&0,  \nonumber \\
\partial _{\phi }\epsilon &=&0,  \nonumber \\
3gq^{I}V_{I} &=&1,  \nonumber \\
-e^{-U}+{\frac{Z}{r}}+grX^{I}V_{I} &=&0,  \nonumber \\
-e^{-U}T^{\prime }-{\frac{Z}{2r^{2}}}+gX^{I}V_{I} &=&0,  \nonumber \\
-e^{-U}V^{\prime }-{\frac{Z}{2r^{2}}}+gX^{I}V_{I} &=&0.  \label{set}
\end{eqnarray}

\bigskip The last two equations in (\ref{set}) implies that one should set $%
T=V.$ Moreover, we make the following choice\footnote{%
one could have set $X^{I}V_{I}$ to an arbitrary constant but the net effect
would be a rescaling of the coupling constant $g$} 
\begin{equation}
X^{I}V_{I}=1,
\end{equation}
Then one immediately obtain from the fifth equation of (\ref{set}), the
following expression for $U$ 
\begin{equation}
e^{-U}={\frac{Z}{r}}+gr,
\end{equation}
Using the last equation of (\ref{set}), we obtain the following differential
equation for $V$, 
\begin{equation}
V^{\prime }e^{-U}=g-\frac{Z}{2r^{2}}.  \label{de}
\end{equation}
The above differential equation can be easily solved by noticing that it can
be rewritten in the following form 
\[
{\frac{dV}{dr}}={\frac{d}{dr}}\log {(e^{-U})}-{\frac{1}{4}}{\frac{d}{dr}}%
\log {({\frac{e^{-U}}{gr}})} 
\]
where we have implicitly assumed that $Z$ takes a constant value to be
determined and therefore one obtains the following solution for $V$ 
\begin{equation}
e^{V}=e^{-3{\frac{U}{4}}}(gr)^{{\frac{1}{4}}}.
\end{equation}

The scalar fields are chosen to minimize the magnetic central charge, as in
the case of the double extreme solutions in the ungauged theory,\cite{feka,
magnetic} i. e., 
\begin{equation}
\partial _{i}Z=\partial _{i}(q^{I}X_{I})={\frac{1}{3}}C_{IJK}X^{I}\partial
_{i}(X^{J})q^{K}=0.  \label{min}
\end{equation}
With the above ansatz, the gaugino transformations given by 
\begin{eqnarray}
\delta \lambda _{i} &=&\left( {\frac{3}{8}}\partial _{i}X_{I}\Gamma ^{\mu
\nu }F_{\mu \nu }^{I}-{\frac{i}{2}}g_{ij}\Gamma ^{\mu }\partial _{\mu }\phi
^{j}+{\frac{3}{2}}igV_{I}\partial _{i}X^{I}\right) \epsilon  \nonumber \\
&=&\left( {\frac{3}{8}}\partial _{i}X_{I}\Gamma ^{\mu \nu }F_{\mu \nu }^{I}+{%
\frac{3i}{4}}\Gamma ^{\mu }\partial _{\mu }X^{I}\partial _{i}X_{I}+{\frac{3}{%
2}}igV_{I}\partial _{i}X^{I}\right) \epsilon
\end{eqnarray}
can be easily seen to vanish identically.

From (\ref{min}), It follows that the critical values of $X^{I}$ and its
dual are given by \cite{magnetic} 
\begin{equation}
X^{I}={\frac{q^{I}}{Z}},\qquad X_{I}={\frac{1}{6}}{\frac{C_{IJK}q^{J}q^{K}}{%
Z^{2}}}
\end{equation}
and thus the critical value of the magnetic central charge is 
\begin{equation}
Z^{3}={\frac{1}{6}}C_{IJK}q^{I}q^{J}q^{K}.
\end{equation}

Using the conditions $X^{I}V_{I}=1$ and the fourth relation of (\ref{set}),
one obtains a generalised Dirac quantization condition 
\begin{equation}
^{3}\sqrt{{\frac{1}{6}}C_{IJK}q^{I}q^{J}q^{K}}={\frac{1}{3g}.}
\end{equation}
The above condition reduces in the case of pure supergravity, i. e, for the
case where there are no vector supermultiplets and one graviphoton charge $%
q^{0}$, to the Dirac quantization of the magnetic charge 
\begin{equation}
q^{0}={\frac{1}{3g}.}
\end{equation}
A similar condition was obtained by \cite{romans} .

\bigskip To summarize, the magnetic string solution to $D=5,$ $N=2$ gauged
supergravity coupled to vector multiplets is given by 
\begin{eqnarray}
ds^{2} &=&(gr)^{\frac{1}{2}}e^{-{\frac{3}{2}}%
U}(-dt^{2}+dz^{2})+e^{2U}dr^{2}+r^{2}\left( d\theta ^{2}+\sin ^{2}\theta
d\phi ^{2}\right)  \nonumber \\
\quad e^{-U} &=&{\frac{1}{3gr}}+gr.
\end{eqnarray}
and the gauge fields and the scalars are given by 
\begin{eqnarray}
A_{\phi }^{I} &=&-q^{I}\cos \theta ,\quad \\
X^{I} &=&3gq^{I}.
\end{eqnarray}

\bigskip The Killing spinor is independent of the angular variables and the
radial dependence of the Killing spinor is obtained by solving for its
radial differential equation given by 
\begin{equation}
\left( \partial _{r}-{\frac{e^{U}}{2}}(g-{\frac{1}{6gr^{2}}})\right)
\epsilon =0
\end{equation}

Using the relation (\ref{de}), the above differential equation can be
written in the following simple form 
\begin{equation}
\left( \partial _{r}-{\frac{1}{2}}V^{\prime }\right) \epsilon =0
\end{equation}
and therefore 
\begin{equation}
\epsilon (r)=e^{{\frac{1}{2}}V}\epsilon _{0}
\end{equation}
where $\epsilon _{0}$ is a constant spinor satisfying the constraints 
\begin{equation}
\Gamma _{3}\Gamma _{4}\epsilon _{0}=i\epsilon _{0},\qquad \Gamma
_{2}\epsilon _{0}=-\epsilon _{0}
\end{equation}

\vfill\eject

\section{Discussions}

In this paper we have obtained explicit magnetic string solutions for all $%
N=2$ supergravity models in five dimensions. The magnetic charges satisfy a
generalised Dirac quantization which for the case of pure supergravity
implies that the magnetic charge is fixed in terms of the inverse of the
coupling constant $g$. A subclass of solutions of $N=2$ supergravity are
also solutions of supergravity theories with $N=4$ and $N=8$
supersymmetries. Those are the gauged versions of the ``toroidal''-type
compactifications.

The fact that the magnetic central charge, for all charge configurations, is
given in terms of the coupling constant $g$ implies that the metric solution
takes a universal form for all gauged theories in five dimensions.

Clearly like the BPS spherical electric solutions in four and five
dimensional gauged supergravity theories, our magnetic string solution
represent a naked singularity. However, it was observed in \cite{klemm} \
that for four dimensional purely magnetic solutions one can get extremal
genuine black holes with event horizons if the two sphere is replaced with
the quotients of the hyperbolic two-space $H^{2}$. In our case, this would
result in the following solitonic solution.

\begin{eqnarray}
ds^{2} &=&(gr)^{\frac{1}{2}}e^{-{\frac{3}{2}}%
U}(-dt^{2}+dz^{2})+e^{2U}dr^{2}+r^{2}\left( d\theta ^{2}+\sinh ^{2}\theta
d\phi ^{2}\right)  \nonumber \\
A_{\phi }^{I} &=&-q^{I}\cosh \theta ,\quad X^{I}=3gq^{I},\quad e^{-U}=gr-{%
\frac{1}{3gr}}
\end{eqnarray}

Details of the above solution and its supersymmetric properties will be
given elsewhere.

$.$

\section*{Acknowledgments}

W. Sabra would like to thank the physics department of Rockefeller
university for hospitality during which some of this work was completed. W.
Sabra would also like to thank Dietmar Klemm for a useful discussion \vfill%
\eject

\end{document}